\documentclass[pre,aps,twocolumn,amsmath,amssymb,longbibliography]{revtex4-1}

\pdfoutput=1
\usepackage{hyperref}
\usepackage{graphicx}

\def\f#1{Fig.~\ref{#1}}

\def\c#1{~\cite{#1}}

\def\f#1{Fig.~\ref{#1}}

\def\s#1{Section~\ref{#1}}

\def\beq{\begin{equation}}
\def\eeq{\end{equation}}
\def\bea{\begin{eqnarray}}
\def\eea{\end{eqnarray}}

\def\kt{k_{\rm B}T}

\begin{document}

\title{Examples of molecular self-assembly at surfaces}

\author{Stephen Whitelam}\email{{\tt swhitelam@lbl.gov}}

\address{Molecular Foundry, Lawrence Berkeley National Laboratory, 1 Cyclotron Road, Berkeley, CA 94720, USA}

\begin{abstract}
The self-assembly of molecules at surfaces can be caused by a range of physical mechanisms. Assembly can be driven by intermolecular forces, or molecule-surface forces, or both; it can result in structures that are in equilibrium or that are kinetically trapped. Here we review examples of self-assembly at surfaces that have been studied within the User program of the Molecular Foundry at Lawrence Berkeley National Laboratory, focusing on a physical understanding of what causes patterns seen in experiment. Some apparently disparate systems can be described in similar physical terms, indicating that simple factors -- such as the geometry and energy scale of intermolecular binding -- are key to understanding the self-assembly of those systems.
\end{abstract}
\maketitle

\section{Assembly by many mechanisms}
\label{intro}

Self-assembly is the spontaneous organization of components into patterns or structures~\cite{whitesides2002self}. Molecular and nanoscale self-assembly is widely studied for its potential for making new materials, but it is also a fascinating example of a fundamental phenomenon, the emergence of order from disorder. Self-assembly at surfaces provides a convenient setting in which to learn more about this phenomenon, in part because it is easier to visualize the dynamics of organization in two dimensions\c{chung2010self} than in three. A large body of work shows that assembly at surfaces can involve a wide range of component and surface types\c{elemans2009molecular,bartels2010tailoring}, and can be result from a range of physical factors. Assembly can happen as a result of molecule-molecule forces\c{he2005self}, or molecule-surface forces, or both\c{otero2011molecular} (see \f{fig1}). Assembly can result in structures that are in equilibrium\c{palma2010atomistic}, or that are kinetically trapped\c{bieri2009porous}, or that have interesting slow dynamical character\c{blunt2008random,otero2008elementary,marschall2010random}. Here we focus on examples of self-assembly at surfaces that have been studied within the Molecular Foundry's User program\c{whitelam2012random,haxton2013competing,whitelam2014common}. These examples highlight the range of physical mechanisms important to self-assembly at surfaces. We show that certain apparently disparate systems can be described in similar physical terms, indicating that simple physical factors -- the geometry and strength of intermolecular binding -- are more directly relevant to the self-assembly of these systems than are the atomic details of the components involved.
\begin{figure*}
   \centering
 \includegraphics[width=\linewidth]{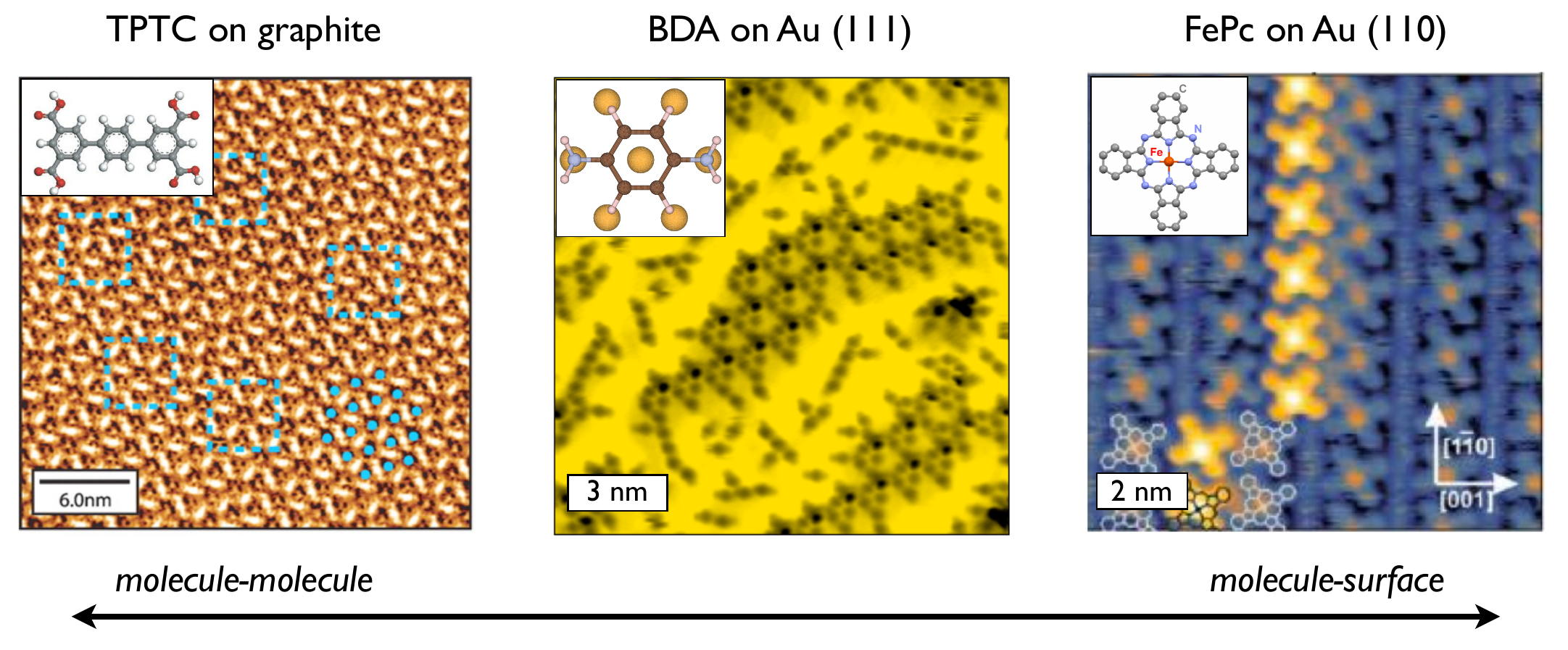} 
   \caption{{\em Self-assembly at surfaces can be driven by molecule-molecule interactions, molecule-surface interactions, or both.} Left: Driven by intermolecular interactions, p-terphenyl-3,5,3Õ,5Õ-tetracarboxylic acid (TPTC) self-assembles into a network equivalent to a random rhombus tiling of the plane\c{blunt2008random} (see \s{sec_rhombus}). Center: On a gold surface, 1,4-substituted benzenediamine (BDA) forms structures whose morphologies are determined by intermolecular interactions, but whose relative abundance depends on molecule-surface interactions\c{haxton2013competing} (see \s{sec_bda}). Right: iron-phthalocyanine (FePc) molecules form chains whose alignment is determined by particular `channels' on the gold surface\c{betti2012structural}. {\footnotesize Left-hand image reproduced from Ref.\c{blunt2008random}, copyright 2008 by The American Association for the Advancement of Science. Center image adapted from Ref.\c{haxton2013competing}. Right-hand image reproduced from Ref.\c{betti2012structural}, copyright (2012) American Chemical Society. Inset to right-hand image reproduced from \href{http://www.ucl.ac.uk/~ucanchi/PyBossa/Molecules/molecules.html}{http://www.ucl.ac.uk/$\sim$ucanchi/PyBossa/Molecules/molecules.html}.}}
\label{fig1}
\end{figure*}

\begin{figure*}[t] 
   \centering
 \includegraphics[width=\linewidth]{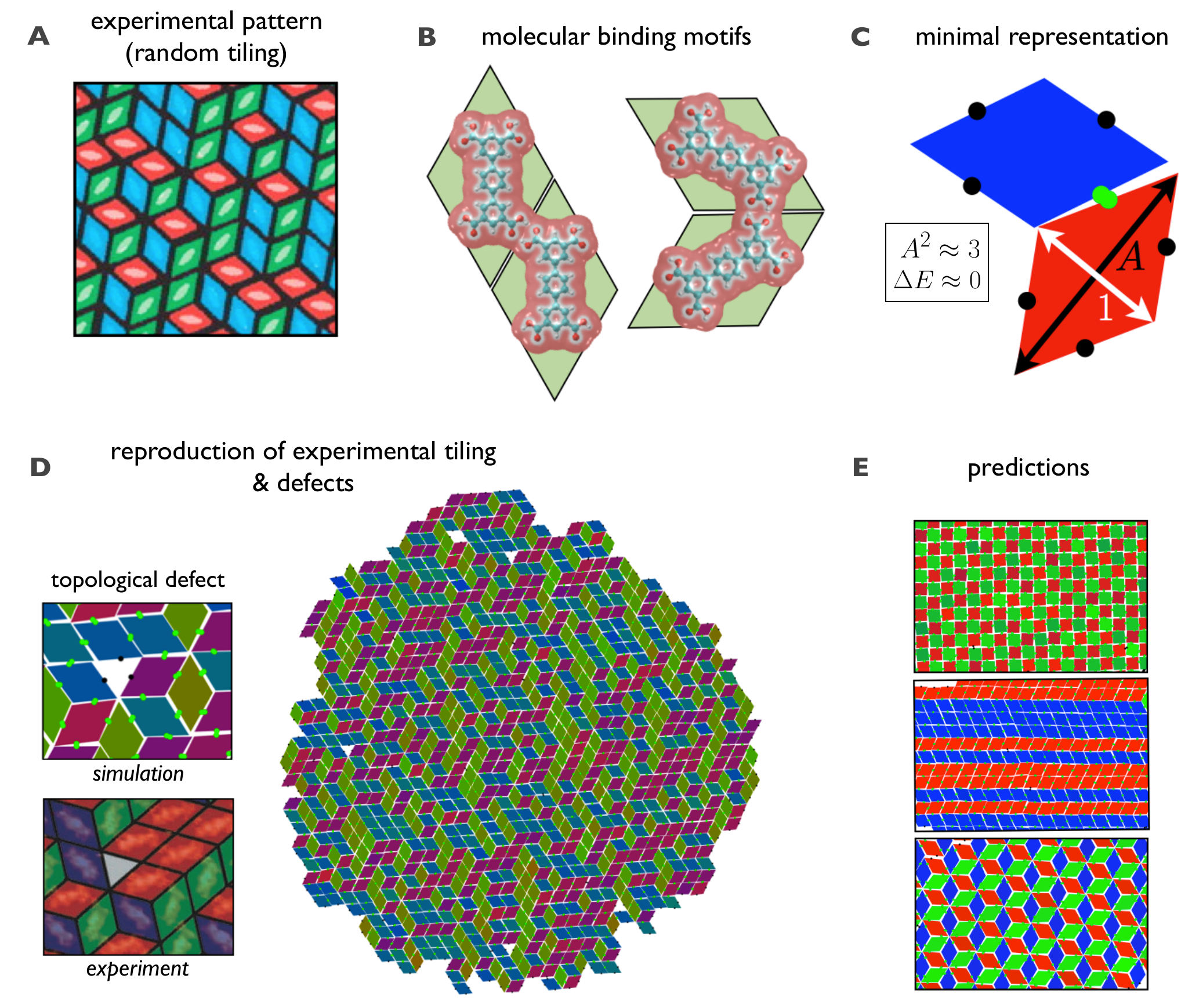} 
   \caption{{\em The self-assembly of TPTC molecules can be reproduced in simulation by accounting only for the geometry and relative energy scales of molecular binding.} (A) TPTC on graphite forms a network structure equivalent to a random rhombus tiling of the plane. In this image, from Ref.\c{blunt2008random}, the bright `rods' are the backbones of the molecules. (B) Quantum mechanical density functional theory calculations show that TPTC engages in pairwise parallel- and 60$^\circ$-rotated hydrogen bonding interactions; these are equal in energy. (C) These two modes of interaction can be reproduced by rhombuses that possess attractive face-to-face interactions and internal angles of 60$^\circ$ and 120$^\circ$. (D) Collections of such rhombuses self-assemble in simulation into the random tiling seen in experiment, including the topological defects that mediate slow rearrangements of tilings. (E) Varying the geometric properties of rhombus tiles leads to alternative self-assembled tilings. {\footnotesize Figure adapted from Ref.\c{whitelam2012random}, except for panel A and the image marked `experiment' in panel D; those images are reproduced from Ref.\c{blunt2008random}, copyright 2008 by The American Association for the Advancement of Science}.}
\label{fig2}
\end{figure*}

\section{Rhombus tilings from rhombus-like interactions}
\label{sec_rhombus}

The molecule p-terphenyl-3,5,3Õ,5Õ-tetracarboxylic acid (TPTC) self-assembles -- following room-temperature deposition from organic solvent onto a graphite surface -- into a network equivalent to a rhombus tiling of the plane\c{blunt2008random}; see \f{fig2}(A). This tiling is stabilized energetically, with respect to a collection of dissociated molecules, by intermolecular hydrogen bonds. The tiling is random, in the sense that it exhibits critical correlations of molecular orientations, and in this respect is stabilized by entropy with respect to other possible ordered rhombus tilings of the plane. 

A combination of quantum mechanics and statistical mechanics can be used to identify the molecular features that cause TPTC to form the random rhombus tiling\c{whitelam2012random}. In \f{fig2}(B) we show the low-energy intermolecular binding motifs of the molecule, as identified by quantum mechanical density functional theory. In these motifs the two molecules are parallel or 60$^\circ$ rotated, and are, to within the confidence limits of the method, equal in energy. The essence of the real intermolecular interactions can be captured using highly simplified building blocks, rhombus tiles that possess sticky patches placed midway along each edge, shown in \f{fig2}(C). Sticky patches mimic hydrogen bonding groups, but tiles otherwise possess no molecular detail. Tiles with internal angles 60$^\circ$ and 120$^\circ$ possess low-free-energy binding motifs that are parallel and 60$^\circ$ rotated, just as the real molecules do. The nucleation and growth of collections of such tiles on a smooth surface results in the random rhombus tiling of the plane, shown in \f{fig2}(D). That assembly in simulations results in structures similar to those seen in experiment shows that the essence of TPTC's self-assembly can be captured by accounting only for a few key aspects of its intermolecular interactions; further atomic detail is not needed. Simulated self-assembled structures also possess the `half-tile' topological defects seen in experimental tilings. These defects mediate slow interconversion of energetically degenerate tilings\c{garrahan2009molecular}, when molecules or tiles adjacent to the defect unbind and are replaced by a molecule or tile at a different angle.

\begin{figure*}[t] 
   \centering
 \includegraphics[width=\linewidth]{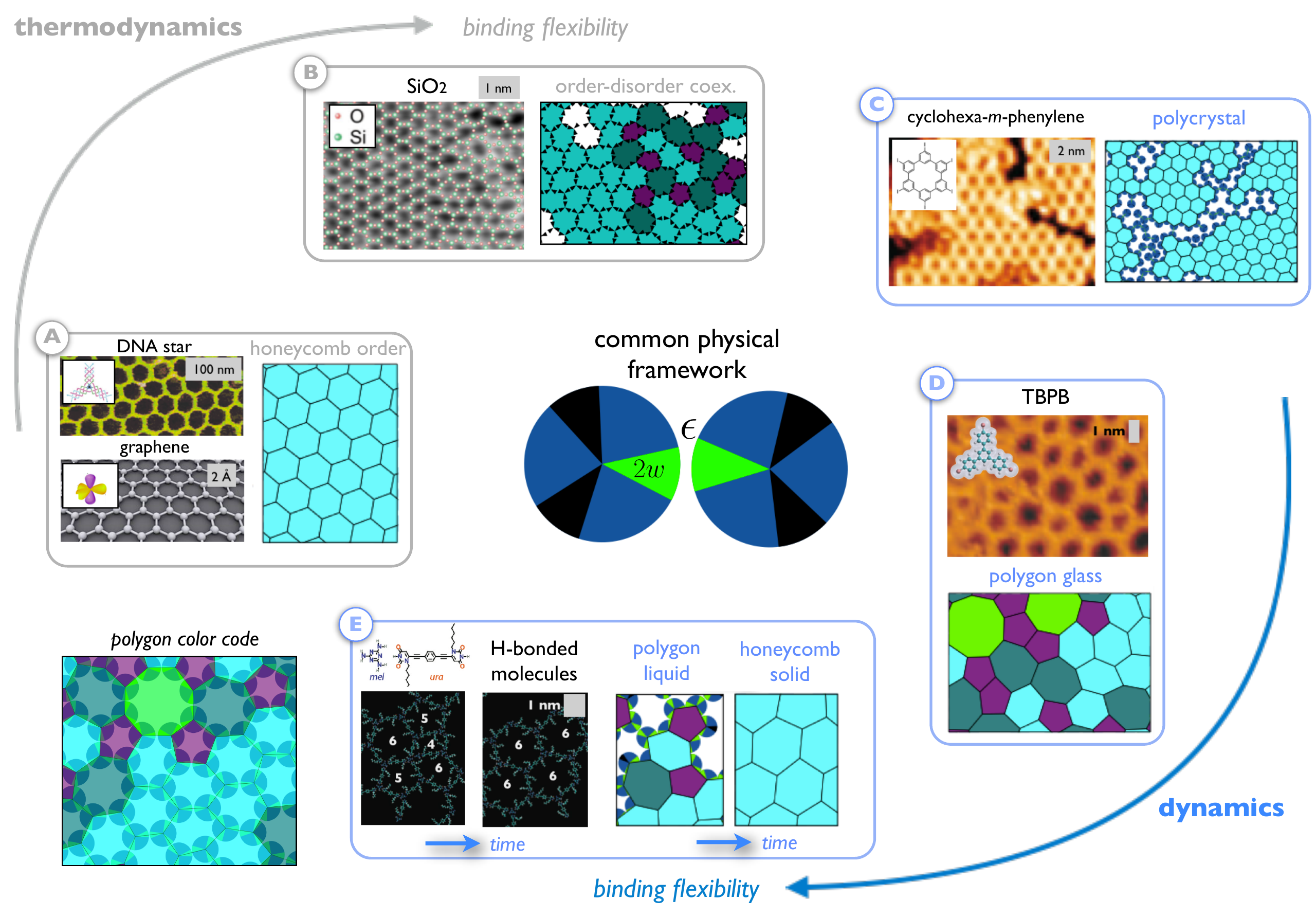} 
   \caption{{\em Qualitative features of the self-assembly of a diverse collection of network-forming components can be reproduced in simulation by accounting only for the geometry and energy scales of molecular binding.} Simulations of a model molecule whose interactions possess regular three-fold rotational symmetry (center) can, upon variation of strength $\epsilon$ and flexibility $w$ of binding, reproduce qualitatively the equilibrium and nonequilibrium behavior of a diverse collection of network-forming components. In equilibrium this behavior includes network order (A), similar to that seen in networks made from carbon atoms\c{geim2007rise} or a DNA star\c{he2005self}; and order-disorder phase coexistence (B), similar to that seen in a silica bilayer\c{lichtenstein2012crystalline}. Away from equilibrium this behavior includes the self-assembly of polycrystals (C), similar to those formed by the small covalently-associating molecule cyclohexa-$m$-phenylene\c{bieri2009porous}; the self-assembly of a kinetically trapped polygon glass (D), similar to that made by the molecule tris(4-bromophenyl)benzene\c{whitelam2014common}; and the formation of a polygon network that evolves to the stable honeycomb one (E), similar to the behavior exhibited by small molecules that associate via hydrogen bonds\c{palma2010atomistic}. Comparison of simulation and experiment suggests that simple design criteria -- here, strength and flexibility of binding -- are more directly relevant to the assembly of these networks than are the precise atomic details of their constituents. Each box (A--E) contains experimental examples together with simulations of discs that show similar features; polygons are drawn atop discs in the manner shown at lower left. {\footnotesize Figure adapted from Ref.\c{whitelam2014common}}.}
\label{fig3}
\end{figure*}

Variation of the geometric and energetic properties of the tile model reveals that molecules that differ even slightly from TPTC would likely not form the random rhombus tiling. A difference in energy between parallel- and nonparallel binding motifs of even small fractions of $\kt$ results in self-assembly of ordered tilings rather than the random one (a fact than can be guessed from classic results of lattice rhombus tilings\c{henley1991random}). Small variations in the aspect ratio of the molecule, or the placement of hydrogen-bonding groups, induce a bias for parallel- or nonparallel binding modes. Nonspecific energetic interactions (e.g. dispersion forces or forces mediated by solvent) can also favor ordered tilings over the random one. In these respects a simulation study can identify physical characteristics of molecules that dispose them to self-assemble in particular ways, and can be used to furnish predictions for the structures that one could build via the self-assembly of other `rhombus-like' molecules; see \f{fig2}(E).

\section{Common description of atomic, molecular and polymeric network-formers}
\label{sec_disc}

The study of TPTC described in the previous section shows that accounting only for coarse features of intermolecular binding geometry and energy scale allows the reproduction of key features of collective behavior seen in experiment. Such a minimal perspective is a common one in statistical mechanics, and many studies have shown that simple models can be used to describe a wide range of systems (see e.g. Refs.\c{molinero2008water,saika2013understanding} for application to tetrahedral liquids). Correspondingly, several examples of molecular self-assembly at surfaces can be described in common physical terms\c{whitelam2014common}. These examples, shown in \f{fig3}, involve components as different as atoms, molecules and polymers, and involve the formation of networks whose basic lengthscales span almost three orders of magnitude. The networks in question are the honeycomb network and its polygon variants. In experiment one sees a range of examples of such networks in equilibrium: carbon atoms\c{geim2007rise} and a DNA star\c{he2005self} form the honeycomb network; a silica bilayer exhibits honeycomb-polygon phase coexistence\c{lichtenstein2012crystalline}. Away from equilibrium one can observe honeycomb polycrystals made from the covalently-associating molecule cyclohexa-$m$-phenylene\c{bieri2009porous}; a polygon glass self-assembled (on gold at about 400 K) from the covalently-associating molecule tris(4-bromophenyl)benzene (TBPB)\c{whitelam2014common}; and an evolving polygon network that relaxes to the honeycomb one, formed from molecules that associate via hydrogen bonds\c{palma2010atomistic}.

What the building blocks (or effective building blocks) of these networks have in common is three-fold coordination that is regular (or nearly so). The simplest model molecule with such properties, a disc with three regularly-spaced sticky patches, is shown in the center of \f{fig3}. A combination of quantum mechanics and statistical mechanics shows that such discs possess the basic polygon-making tendencies of one of these experimental systems, the molecule TBPB. When arranged into isolated polygons, both discs and molecules favor the hexagon, but they can with some free-energy cost form other polygons. This cost can be controlled, within the disc model, by varying the width of the sticky patch, and can be made similar to that of TBPB. Polygons are the key features of the experimental networks, and so the disc-molecule polygon-making correspondence suggests that assembly of discs into polygon networks should possess some of the characteristics of assembly of the experimental system. This is indeed the case [\f{fig3}(D)]. Moreover, the disc model can reproduce the self-assembly behavior of the other experimental systems, indicating that those systems possess a common physical underpinning [\f{fig3}(A--E)].

This behavior is as follows. In equilibrium, the disc model undergoes a phase transition between network order and disorder. Network order (the honeycomb) is favored by rotational entropy, because discs within hexagons possess more rotational entropy than do discs within other polygons. Network disorder (polygon networks) is favored by configurational entropy. The relative importance of these factors depends on disc binding flexibility, or patch width, in terms of which a phase transition between network order and disorder occurs (similar physics was previously observed in 3D networks\c{smallenburg2013liquids}). The disc model therefore displays the range of equilibrium behavior -- interpolating between network order and disorder -- that is seen in experiment\c{geim2007rise,he2005self,lichtenstein2012crystalline}. 
\label{sec_bda}
\begin{figure*}[t] 
   \centering
 \includegraphics[width=\linewidth]{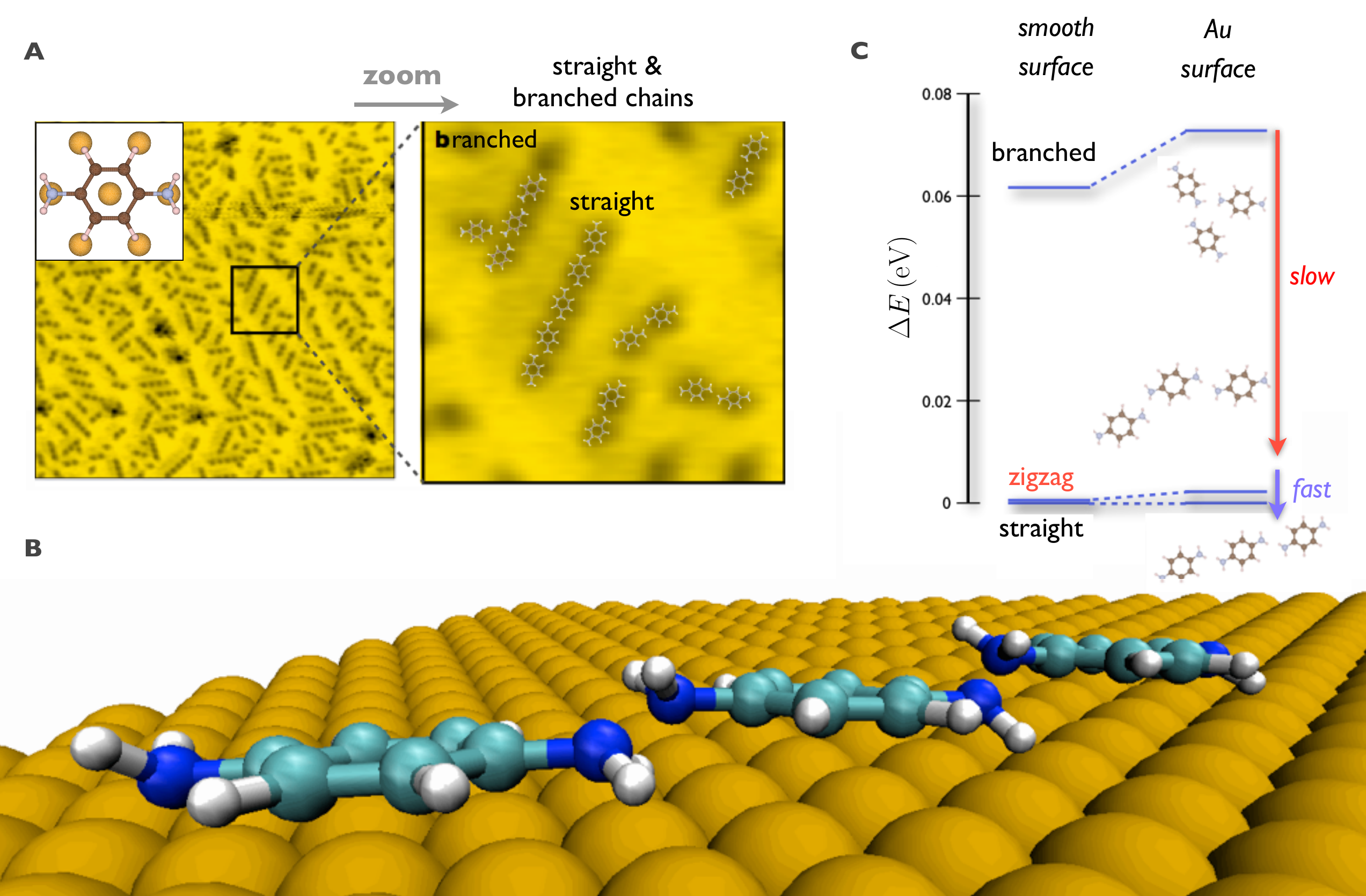} 
   \caption{{\em Patterns seen in experiment can result from a competition of microscopic forces and of thermodynamic and dynamic factors.} (A) BDA on gold forms branched and straight chains, and not zigzag chains (which in vacuum are equal in energy to straight chains). (B) Analysis of a statistical mechanical model that is parameterized using quantum mechanics suggests (C) that these motifs result from intermolecular hydrogen bonding, and that the surface favors straight chains over zigzag ones. Branched chains, higher in energy than both of those chain types, are seen because they are kinetically trapped. {\footnotesize Figure adapted from Ref.\c{haxton2013competing}}.}
\label{fig4}
\end{figure*}

Out of equilibrium, discs self-assemble via dynamical pathways that change qualitatively upon variation of intermolecular binding strength and energy scale. Discs with strong, inflexible interactions form honeycomb polycrystals, similar to cyclohexa-$m$-phenylene\c{bieri2009porous}. Because interactions are inflexible, discs form hexagons much more readily than they form other polygons. Because interactions are strong, nucleation is rapid, and incommensurate grains meet and form disordered boundaries that fail to relax in the time of observation. Discs with strong, flexible interactions form polygon glasses, similar to TBPB. Because interactions are flexible, discs associate initially into polygons of a range of types. Because bonds are strong (note that the energy scale of TBPB's covalent bonds is about 150 $\kt$ at experimental temperatures), these polygons cannot evolve further, and the network fails to relax to the thermodynamically stable honeycomb one. The statistics of polygon networks in this regime depends on preparation protocol, including rates of cooling and relative rates of molecular deposition and on-surface diffusion. Discs with weak, flexible interactions form a polygon network that evolves to the thermodynamically stable honeycomb one, similar to the reversibly-associating molecules of Ref.\c{palma2010atomistic}. Inflexible interactions allow the association of discs into polygons of a range of types; weak interactions allow the frequent breaking and making of intermolecular bonds, a process that permits evolution to the stable honeycomb network. 

Thus, a range of behavior seen in a variety of experimental systems can be reproduced by varying two parameters of a minimal model. These parameters, related to the flexibility and energy scale of molecular binding, can then be regarded as the key physical characteristics of the experimental systems in respect of their self-assembly. The atomic details of the components in question are important insofar as they give rise to these characteristics, but materials that look different in an atomic sense can be modeled in common terms. Predictions for new materials can then be made by drawing on these commonalities: for instance, it is natural to predict that colloids with three-fold-rotational symmetry could self-assemble into a kinetically trapped polygon network, if their interactions are made sufficiently flexible and strong. Such a material would have unusual photonic properties\c{florescu2009designer}.

\section{Pattern selection by competing mechanisms}

So far we have focused on examples of self-assembly at surfaces in which the surface plays host to the molecular overlayer but does not strongly influence it. But there are many examples in which it does\c{otero2011molecular}, one of which is 1,4-substituted benzenediamine (BDA) on gold\c{haxton2013competing}. \f{fig4}(A) shows STM images taken after gas-phase deposition of BDA onto gold, during which the system was cooled to 5 K from about 300 K. In this and in similar images, hundreds of straight chains and branched chains of BDA molecules are seen. Quantum mechanical density functional theory shows that these motifs are low-energy structures stabilized by intermolecular hydrogen bonds. But DFT also reveals the existence of a puzzle: a zigzag variant of the straight chain is in vacuum energetically equivalent to it (essentially by symmetry), but no zigzag chains are seen in STM images.

Resolving this puzzle requires an assessment of the effect of the surface, and of thermodynamic and dynamic factors, which can be done by building a statistical mechanical model of BDA on gold, parameterized using DFT [\f{fig4}(B)]. Model calculations show that at temperatures below about 100 K the surface begins to exert a thermodynamic preference for straight chains over zigzag ones. This observation suggests that zigzag chains are absent from images because they convert to straight chains, preferred energetically by the substrate, as temperature is reduced. However, branched chains are considerably higher in energy than both straight and zigzag chains, and yet are plentiful in images. Dynamic calculations suggest that branched chains are seen because they are kinetically trapped: once formed, they cannot convert to straight chains before the system is cooled to 5 K, where nanostructures are essentially immobile. 

Thus, intermolecular interactions stabilize three low-energy chain types: straight and zigzag, equally; and branched, to a lesser degree. On a cold gold surface, straight chains are preferred thermodynamically to zigzag ones, and both are preferred to branched ones. And while the relaxation dynamics of zigzag chains is fast enough to allow them to convert to straight chains during cooling, branched chains relax too sluggishly to do so. As a consequence, STM images contain a mixture of low-free-energy structures (straight chains); kinetically trapped high-free-energy structures (branched chains); but do not contain intermediate free-energy structures (zigzag chains). The apparently simple patterns seen in images arise through a competition of intermolecular and surface-mediated forces, thermodynamic and dynamic factors, and so highlight the complexity that can arise in self-assembly at surfaces. 

A major cause of this complexity is the fact that the system possesses a distribution of energy scales, and at low temperature much of this distribution matters. For instance, at low temperature the energetic modulation felt by molecules on the surface becomes large enough, relative to $\kt$, to influence the patterns that molecules form (by contrast, many of the experiments described in \s{sec_rhombus} and \s{sec_disc} were done at room temperature or above). Moreover, at low temperature one also encounters significant obstacles to theory and simulation. At 5 K, small energetic differences between molecular configurations -- e.g. on the meV scale -- exceed the thermal energy $\kt$, and so matter significantly to the thermodynamics of assembly. But it is difficult to calculate energetic differences to this degree of precision: DFT is not generally accurate to better than meV. In addition, at low temperature the timescale for significant motion of nanostructures is so long as to make dynamical simulation of even simplified models impractical. Thus, while a combination of quantum mechanics and statistical mechanics can give physical insight into the factors that select patterns of BDA molecules seen in STM images, quantitative prediction of such patterns, at low temperature, is probably beyond current methods.

\section{Outlook}

Molecular self-assembly at surfaces has practical application, and is of fundamental interest. By studying it, we learn about the physical factors that induce components to undergo a particular dynamics of organization. Under some conditions, particularly at low temperature, intermolecular and molecule-surface forces compete to select patterns (\s{sec_bda}). In other settings, evidence suggests that a range of systems can be described in similar physical terms (Sections \ref{sec_rhombus} and \ref{sec_disc}). This similarity implies the existence of organizing principles that, if better understood, could aid the development of new materials. 

\section{Acknowledgements}
This paper was prepared as a Research News article for a forthcoming special issue of Advanced Materials that will focus on work done at the Molecular Foundry at Lawrence Berkeley National Laboratory. This paper summarizes some recent User projects done at the Molecular Foundry, Lawrence Berkeley National Laboratory, supported by the Office of Science, Office of Basic Energy Sciences, of the U.S. Department of Energy under Contract No. DE-AC02--05CH11231. These projects resulted in Refs.\c{whitelam2012random,haxton2013competing,whitelam2014common}, and I would like to thank my co-authors on those papers for enjoyable collaborations.

\onecolumngrid


%

\clearpage

\end{document}